%% file: main.tex
\documentclass[runningheads]{llncs}
\usepackage[utf8]{inputenc}
\usepackage[T1]{fontenc}
\usepackage{graphicx}
\graphicspath{ {./images/} }

\input{addon/pkgloading}%
\usepackage{times}

\input{addon/glossary}
\input{addon/glossary-BPM}

\input{addon/macros}
\begin{document}
\title{Fine-grained Data Access Control for Collaborative Process Execution on Blockchain}
\titlerunning{Access Control on Blockchain}
%
\author{Edoardo~Marangone\inst{1}\orcidID{0000-0002-0565-9168} 
	\and
	Claudio~Di~Ciccio\inst{1}\orcidID{0000-0001-5570-0475} 
	\and
	Ingo~Weber\inst{2}\orcidID{0000-0002-4833-5921}
}
\authorrunning{E.\ Marangone et al.}
%
\institute{Sapienza University of Rome, Rome, Italy\\
	\email{\href{mailto:edoardo.marangone@uniroma1.it;claudio.diciccio@uniroma1.it}{\{edoardo.marangone,claudio.diciccio\}@uniroma1.it}}
	\and
	Software and Business Engineering, Technische Universitaet Berlin, Germany\\
	\email{\href{mailto:ingo.weber@tu-berlin.de}{ingo.weber@tu-berlin.de}}
}
\maketitle              
\begin{abstract}
	\input{sections/abstract}
\end{abstract}

\section{Introduction}
\label{sec:intro}
\input{sections/intro}

\section{Running example and problem illustration}
\label{sec:example}
\input{sections/example}

\section{Background}
\label{sec:background}
\input{sections/background}

%
\label{sec:abe}
\input{sections/abe}

\section{The CAKE approach}
\label{sec:approach}
\input{sections/approach}

\section{Implementation and evaluation}
\label{sec:imptes}
\input{sections/imptes}

\section{Related work}
\label{sec:sota}
\input{sections/sota}

\section{Conclusion and future remarks}
\label{sec:conclusion}
\input{sections/conclusion}

\medskip
\noindent\textbf{Acknowledgements.} The work of E.~Marangone and C.~Di~Ciccio was partially funded by the Cyber 4.0 project BRIE and by the Sapienza research projects SPECTRA and ``Drones as a service for first emergency response''.

%
%
\bibliographystyle{splncs04}
\bibliography{bibliography}

\end{document}

%% file: addon/pkgloading.tex
\usepackage[english]{babel}
\usepackage{subcaption}
\usepackage{url}
\usepackage{colortbl}
\usepackage{amssymb}
\usepackage{stmaryrd}
\usepackage{amsmath}
\usepackage{mathrsfs}
\usepackage{amssymb}
\usepackage[left]{lineno}
\usepackage{xfrac}
\usepackage{nicefrac}
\usepackage[textsize=scriptsize,backgroundcolor=yellow!40]{todonotes}
\usepackage[nonumberlist,acronym,sanitize=none]{glossaries}
\glsdisablehyper
\usepackage{comment}
\usepackage[pdftex, colorlinks=true, hyperfootnotes=true, hyperindex=true,
            plainpages=false, pagebackref=false, pdfpagelabels=true, pdfstartview=FitH,
            linkcolor=blue, citecolor=blue, urlcolor=blue,
            bookmarks, bookmarksopen, bookmarksdepth=3]{hyperref}
\usepackage[capitalise,nameinlink]{cleveref}
\crefname{algocf}{alg.}{algs.}
\Crefname{algocf}{Algorithm}{Algorithms}
\crefname{section}{Sect.}{Sects.}
\usepackage{tkz-base}
\usetikzlibrary{decorations.pathmorphing,trees,snakes,arrows,shapes,automata,petri}
\usepackage{paralist}
\usepackage{multicol}
\usepackage{booktabs}
\usepackage{enumitem}
\usepackage{multirow}
\usepackage{rotating}
\usepackage{etaremune}
\usepackage{marginnote}
\usepackage{mathtools}
\usepackage{mathabx}
\usepackage{adjustbox}
\usepackage{ifthen}
\usepackage[normalem]{ulem}
\usepackage{lineno}
\usepackage{soul}
\usepackage{floatrow}
\usepackage{listings}
\crefname{lstlisting}{Listing}{Listings}
\usepackage{lipsum}
\usepackage{makecell}
\usepackage{diagbox}
\usepackage[scientific-notation=false]{siunitx}
\usepackage{microtype}
\usepackage{footmisc}
\usepackage{xspace}

%% file: addon/glossary.tex
\newacronym[\glslongpluralkey={Distributed Ledger Technologies}]{dlt}{DLT}{Distributed Ledger Technology}
\newcommand{\DLT}[0] {\Gls{dlt}\xspace}
\newcommand{\DLTs}[0] {\Glspl{dlt}\xspace}

\newacronym{ipfs}{IPFS}{InterPlanetary File System}
\newcommand{\IPFS}[0] {\Gls{ipfs}\xspace}

\newacronym{p2p}{P2P}{peer-to-peer}
\newcommand{\PtoP}[0] {\Gls{p2p}\xspace}

\newacronym{abe}{ABE}{Attribute-Based Encryption}
\newcommand{\ABE}[0] {\Gls{abe}\xspace}

\newacronym{ssl}{SSL}{Secure Sockets Layer}

\newacronym{dht}{DHT}{Distributed Hash Table}
\newcommand{\DHT}[0] {\Gls{dht}\xspace}

%% file: addon/glossary-BPM.tex
\newacronym[\glslongpluralkey={Business Processes}]{bp}{BP}{Business Process}
\newacronym{bpi}{BPI}{Business Process Intelligence}
\newacronym{bpm}{BPM}{Business Process Management}
\newacronym{bpms}{BPMS}{Business Process Management System}
\newacronym{bpmn}{BPMN}{Business Process Model and Notation}
\newacronym{cpn}{CPN}{colored Petri net}
\newacronym{kpi}{KPI}{Key Performance Indicator}
\newacronym{ocbc}{OCBC}{Object-centric Behavioral Constraints}
\newacronym{soa}{SOA}{Service-Oriented Architecture}
\newacronym{pn}{PN}{Petri net}
\newacronym{wf}{WF}{workflow}
\newacronym{wfms}{WfMS}{Workflow Management System}
\newacronym{xes}{XES}{eXtensible Event Stream}
\newacronym{yawl}{YAWL}{Yet Another Workflow Language}
\newglossaryentry{task}{%
	name={task},description={the non-divisible, elementary activity}}

\newglossaryentry{promod}{%
	name={process model},description={the model of a process}
}
\def\LogAlph {\ensuremath{\Sigma}}
\newglossaryentry{logalph}{
	name={log alphabet},description={the process alphabet, as reflected in a log},%
	symbol={\LogAlph}}
\def\Evt {\ensuremath{e}}
\newglossaryentry{evt}{
	name={event},description={a record of an instantaneous fact during the process enactment},%
	symbol={\Evt}}
\def\Trc { \ensuremath{\tau} }

\newglossaryentry{trace}{
	name={trace},description={a sequence of \glsplural{evt}},%
	symbol={\Trc}}
\def\EvtLog {\ensuremath{L}}

\newglossaryentry{evtlog}{
	name={event log},description={a collection of \glstext{evttrace}s},%
	symbol={\EvtLog}}

%% file: addon/macros.tex
\renewcommand\tabcolsep{5pt}
\newcolumntype{d}{>{\columncolor{gray!10}}c}
\newcolumntype{m}{>{\columncolor{gray!10}}l}
\newfloatcommand{capbtabbox}{table}[][\FBwidth]
\newenvironment{iiilist}%
{\begin{inparaenum}[\itshape(i)\upshape]}%
{\end{inparaenum}}

\RequirePackage{xparse}
\NewDocumentEnvironment{AuthNote}{+o+o}{%
	\IfValueT{#2}{\marginnote{\scriptsize{#2}}}%
	\begin{scriptsize}
		\colorbox{gray}%
		{\color{white} Note\IfValueT{#1}{ (#1)}:}%
		\quad%
		\color{brown}
}{%
	\normalcolor
	\end{scriptsize}
}

\RequirePackage{pifont}

\RequirePackage{lipsum}
\newcommand{\LipsumGray}[1][]{{\color{gray}\ifthenelse{\equal{#1}{}}{\lipsum}{\lipsum[#1]}}}

\RequirePackage{siunitx}
\newcolumntype{D}[1]{S[
	table-omit-exponent,
	round-mode=places,
	round-integer-to-decimal,
	round-precision={#1}]} 

%% file: sections/abstract.tex
Multi-party business processes are based on the cooperation of different actors in a distributed setting. Blockchains can provide support for the automation of such processes, even in conditions of partial trust among the participants. On-chain data are stored in all replicas of the ledger and therefore accessible to all nodes that are in the network. Although this fosters traceability, integrity, and persistence, it undermines the adoption of public blockchains for process automation since it conflicts with typical confidentiality requirements in enterprise settings. In this paper, we propose a novel approach and software architecture that allow for fine-grained access control over process data on the level of parts of messages. In our approach, encrypted data are stored in a distributed space linked to the blockchain system backing the process execution; data owners specify access policies to control which users can read which parts of the information. To achieve the desired properties, we utilise Attribute-Based Encryption for the storage of data, and smart contracts for access control, integrity, and linking to process data. We implemented the approach in a proof-of-concept and conduct a case study in supply-chain management. From the experiments, we find our architecture to be robust while still keeping execution costs reasonably low.

%% file: sections/intro.tex
Blockchain technology is gaining momentum, among other reasons because it allows for the creation and enactment of business processes between multiple parties with low mutual trust~\cite{Weber.etal/BPM2016:UntrustedBusinessProcessMonitoringandExecutionUsingBlockchain,Stiehle22SLR}. 
The distributed nature of public permissionless blockchains allows every user in the network to have a copy of the ledger and therefore all the data is freely accessible.
This transparency, together with the permanence of data and non-repudiability of transactions granted by the technology, motivate the use of blockchains as a reliable ground for verifiable and trustworthy interactions.

Especially in cases wherein the parties lack trust in one another, though, hiding some data from the majority of users can be useful. As a matter of fact, security and privacy are at the centre of the debate when considering blockchain technology~\cite{Privacy1,Privacy2}.
For example, Corradini et al.~\cite{Corradini.etal/ACMTMIS2022:EngineeringChoreographyBlockchain} point to security and privacy aspects as relevant points. 
The authors underline that the encryption of the payload of messages (a solution already present in the literature) does not preserve the secrecy of information.
Sharing a decryption key among process participants does not allow data owners to selectively control the access to different parts of a single message. Using the public key of a recipient forces the sender to create multiple copies of every message (one per intended reader) and severely hampers the traceability of the process.
Another proposed solution is the usage of permissioned blockchains. However, this scheme entails strong complexity and management issues.

Our work aims to close the gap by proposing a technique that guarantees data privacy among parties. With this architecture, the parties can exchange information in a secure way and can also hide data (or parts thereof) from other players with whom they do not want to share it. As such, this paper introduces a novel approach to address security and privacy problems by presenting an architecture that allows for the ciphering of selected data using Attribute-Based Encryption (ABE)~\cite{ABE} so as to control fine-grained read and write access to data.


In the following, \cref{sec:example} presents a running example, to which we will refer throughout the paper, and illustrates the problem we tackle. \Cref{sec:background} outlines the fundamental notions that our solution is based upon. In \cref{sec:approach}, we describe our approach in detail. In \cref{sec:imptes}, we present our proof-of-concept implementation and illustrate the results of the experiments we conducted therewith. \Cref{sec:sota} presents the related work in the literature. Finally, \cref{sec:conclusion} concludes the paper and draws some avenues for future works.

%% file: sections/example.tex
\begin{figure}[tb]
  \includegraphics[width=\textwidth]{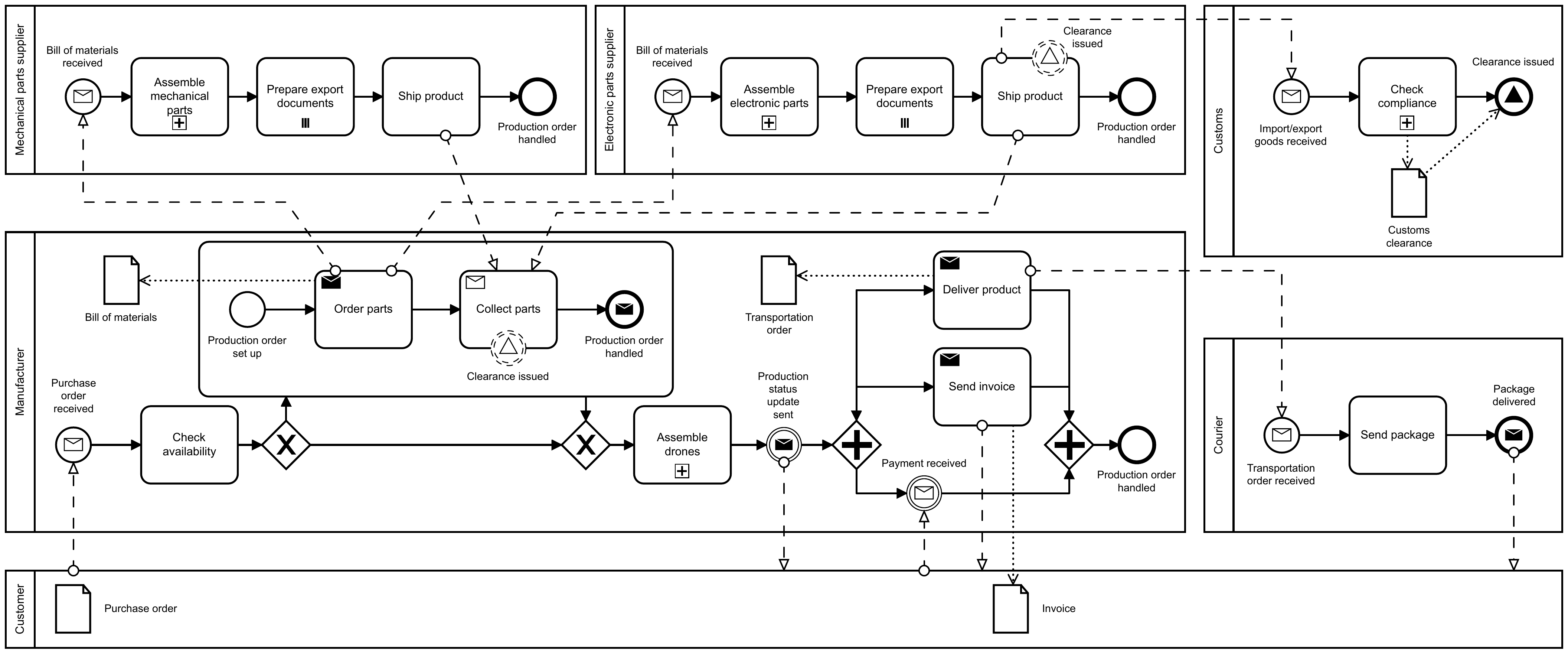}
  \caption{BPMN collaboration diagram of a multi-party process}
  \label{fig:example}
\end{figure}

\Cref{fig:example} depicts a \gls{bpmn} collaboration diagram representing the supply chain behind the production of drones.
We will use this scenario as a running example throughout our paper. 
We assume process execution is backed by a blockchain-based infrastructure as illustrated in~\cite{DiCiccio.etal/InfSpektrum2019:BlockchainSupportforCollaborativeBusinessProcesses}. 
%

A new process instance begins when a \textit{Customer} orders one or multiple drones from a \textit{Manufacturer}. 
After checking the availability of the mechanical and electronic components in the warehouse, the \textit{Manufacturer} orders the missing ones from a local \textit{Mechanical parts supplier} and an international \textit{Electronic parts supplier}, respectively. After the assemblage of the required parts, the suppliers prepare the shipment documents, the package, and send the products.
\textit{Customs} then check the documents of the international supplier and release the custom clearance after the verification of compliance concludes positively.
Upon the receipt of the parts, the \textit{Manufacturer} proceeds with their assemblage. After sending a notification about the stage reached by the production process, the \textit{Manufacturer} sends an invoice to the paying \textit{Customer}, and requests a \textit{Courier} to deliver the package. The process concludes with the consignment of the ordered product.

We highlight the information artefacts we are going to primarily focus on for our examples as paper documents with twisted corners, namely
\begin{iiilist}
	\item purchase order, 
	\item bill of materials (BoM), 
	\item customs clearance, 
	\item invoice (for the customer), and
	\item transportation order.
\end{iiilist}
First and foremost, we observe that the exchanged information in this process should not be fully accessible outside of the involved counterparts in the process execution.
Notice that, instead, a non-encrypted communication through the blockchain allows every node (not necessarily involved in the process either) to disclose the full content of all data attached to transactions. 
If all parties knew a secret key, they could store the data on-chain once encrypted with that key to ensure nobody outside their circle can read through them, yet ensuring that the data are notarised by the blockchain.
However, we remark that although the information exchanges involve multiple actors in collaborative processes, it is rare that \emph{every} actor is supposed to read \emph{all} the exchanged data in their entirety -- particularly in this scenario, it never is the case.
For example, the invoice details should be undisclosed to any other party that is not the \textit{Customer} or the \textit{Manufacturer}, just like the purchase order.
Likewise, the transportation order should be fully accessible only to the \textit{Manufacturer} and the \textit{Courier}.

Whenever a message sender and recipient are single players who know one another in advance, the data producer could encrypt the message and give the access key to the sole expected consumer.
However, this may not be a reasonable assumption in cases like the one we discuss here.
The customs clearance, for instance, should be known to more than two parties, as the \textit{Electronic parts supplier} and the \textit{Customs} are directly involved but the \textit{Manufacturer} should also be made aware of the result at the end of the border controls.
Besides, not only the operators in the \textit{Customs} office involved in the first inspection should be granted access -- this restriction would impede future checks.

Another example of non-binary communication channel pertains to the bill of materials. The section of the BoM for the \textit{Mechanical parts supplier} should not be read by any other party but the recipient of the production order and the \textit{Manufacturer}.
Notice that, albeit the \textit{Electronic parts supplier} is also a producer of basic components for the \textit{Manufacturer}, it should access the sole part of the BoM referred to its area of competence.
Therefore, different parts of a shared data artefact should be accessible to different players.
In contrast, the section of the BoM with the identifying data of the \textit{Manufacturer} should be visible to both suppliers.

In the last few years, research work flourished for blockchain-based control-flow automation and decision support for processes like the one in this section ~\cite{Weber.etal/BPM2016:UntrustedBusinessProcessMonitoringandExecutionUsingBlockchain,Tran.etal/BPMDemos2018:Lorikeet,Lopez-Pintado.etal/SPE2019:Caterpillar,Madsen.etal/FAB2018:CollaborationamongAdversaries:DistributedWorkflowExecutiononaBlockchain}. Our investigation complements this body of research by focussing on the 
secure information exchange among multiple parties in a collaborative though partially untrusted scenario. 
We list the key requirements for our approach in \cref{tab:requirements}.
Next, we focus on the background knowledge that to which our approach resorts.

\begin{table}[tb]
	\centering
	\begin{tabular}{p{0.5cm} p{4cm} p{6.5cm}} \toprule
		& \textbf{Requirement} & \textbf{Approach} \\ \midrule
		\textbf{R1} & Access to parts of messages should be controllable in a fine-grained way (attribute level), while integrity is ensured &  We use Attribute-Based Encryption (ABE) to encrypt messages, which are stored off-chain while their locator and hash are kept on chain. Access is mediated by a component that decrypts messages only if the requester has the necessary attributes. \\
		\textbf{R1.1} & Access policies should be linked to individual (parts of) information artefacts & Access policies associate granted classes of users to the sole messages or sections (slices) thereof that pertain to them \\
		\textbf{R1.2} & Access policies should control access levels for authenticated users & The policies are fine-grained, and the component that decrypts messages does so as per on-chain information \\
		\textbf{R1.3} & Non-authorised access is prevented & Data is kept in an encrypted form, and only authorised requests allow for decryption; salting prevents leakage of information through hashes \\
		\textbf{R2} & Information artefacts should be written in a permanent, tamper-proof and non-repudiable way & We use hashed, permanent off-chain storage in combination with hashes on-chain \\
		\textbf{R3} & The system should be independently auditable with low overhead & On-chain information is publicly available to users of the system, and through hashes integrity of off-chain data becomes auditable \\
		\bottomrule
	\end{tabular}
	\caption{Requirements and corresponding actions in the approach}
	\label{tab:requirements}
\end{table}

%% file: sections/background.tex
In this section, we give an overview of the fundamental notions upon which our approach is built. The fundamental building blocks of our work are \acrfull{dlt}, particularly programmable blockchain platforms, and \acrfull{abe}.
Next, we outline the basic notions they build upon and relate them to our running example.

\DLTs
realise protocols for the storage, processing and validation of transactions among a network of peers.
Their distributed nature entails that no central authority or intermediary are involved in the management of the data.
To all these transactions a timestamp and a unique cryptographic signature are attached.
To produce signatures, a public/private key scheme is adopted. Every user holds an account with a unique address to which the public and private keys are associated.
The shared database is public so all 
participants in the network can 
have access to the data. 
\textbf{Blockchain}
is a type of \DLT, wherein segments of the ledger are collated into blocks and those blocks are backward-linked together forming a chain. 
\DLTs in general and the blockchain in particular cannot be tampered with thanks to a blend of cryptographic techniques, including the hashing of blocks themselves, the inclusion in every block of the hash of the previous one, and the distributed validation of transactions. 
%
Public blockchain platforms such as Bitcoin~\cite{Nakamoto/2008:Bitcoin:APeer-to-PeerElectronicCashSystem}, Ethereum~\cite{Wood/2018:Ethereum} and Algorand~\cite{Chen.Micali/TCS2019:Algorand} require fees to be paid in order to let transactions be submitted and processed by the platform.
More recent blockchain protocols such as Ethereum and Algorand include the opportunity to run \textbf{Smart Contracts}, namely
programs deployed, stored and executed in the blockchain~\cite{Dannen/2017:IntroducingEthereumandSolidity}. 
Smart contracts are invoked via transactions.
The execution is spread among the nodes without the involvement of a trusted third party so that the overall behaviour can be verified and trusted. Moreover, smart contracts can also trigger the next steps of a workflow when some conditions are met~\cite{Weber.etal/BPM2016:UntrustedBusinessProcessMonitoringandExecutionUsingBlockchain}.
As with transactions, the execution of smart contract code is subject to costs that in the Ethereum nomenclature fall under the name of \emph{gas}.
These costs depend on the complexity of the invoked code and on the amount of data exchanged and stored.
To reduce the invocation costs of smart contracts, external \PtoP systems are typically employed to save larger bulks of data~\cite{Xu.etal/2019:ArchitectureforBlockchainApplications}. One such system is the \textbf{\IPFS}, a distributed system for 
the storage and access to files.
Having it a \DHT at its core, the stored files are scattered among several nodes. 
Akin to \DLTs, no central authority or trusted organisation retaining the whole bulk of data is thus involved. \IPFS makes use of content-addressing to uniquely identify each file in the network. 
The data saved on \IPFS are hash-linked by resource locators that are then sent to contracts that store them on chain~\cite{Lopez-Pintado.etal/IS2022:ControlledFlexibilityBlockchainCollaborativeProcesses}. Thereby, the hash of external data together with their remote handle are permanently stored on chain to link them to the ledger.

In a multi-party collaboration scenario like the one we described in \cref{sec:example}, the blockchain creates a layer of trust: the ledger operates as an auditable notarisation infrastructure to certify the occurrence of transactions among the involved actors (e.g., the purchase orders or custom clearances), the smart contracts guarantee that the workflow is followed as per the agreed behaviour, as illustrated in \cite{DiCiccio.etal/InfSpektrum2019:BlockchainSupportforCollaborativeBusinessProcesses,Mendling.etal/ACMTMIS2018:BlockchainsforBPM}.
Documents such as purchase orders, bill of materials and custom clearances can be stored on \IPFS and linked to the transactions that report on their submission.
However, those data are accessible to all peers on chain.
Techniques to cipher the data and control their accessibility to predefined users become necessary so as to take advantage of the security and traceability guarantees of blockchain while managing read and write grants on the stored information.

%% file: sections/abe.tex
\textbf{\ABE} is a type of public-key encryption in which the \emph{ciphertext} (i.e., an encrypted text derived from a \emph{plaintext}) and the corresponding private key to decipher it are linked together by attributes~\cite{ABE}.
In particular, the Ciphertext-Policy \ABE (CP~\cite{CP-ABE}) is such that every potential user is associated with a number of \emph{attributes} over which \emph{policies} are expressed.
Attributes, in particular, are propositional literals that are affirmed in case a user enjoys a given property.
In the following, we shall use the teletype font to format attributes and policies.
For example, user \texttt{0xE756[...]b927} is associated with the attributes \texttt{Supplier}, to denote their role, and  \texttt{14548487}, to specify their involvement in process instance number \texttt{14548487}. For the sake of brevity, we omit from the attribute name that the former is a role and the latter a process instance identifier (e.g., \texttt{Supplier} in place of \texttt{RoleIsSupplier} or \texttt{14548487} instead of \texttt{InvolvedIn14548487}) as we assume it is understandable from the context.
Policies are associated to messages and expressed as propositional formulae on the attributes (the literals) to determine whether a user is granted access (e.g., \texttt{Courier or Manufacturer}).

All users can attain a unique \emph{secret key} (\textsl{sk}).
The \textsl{sk} is a fixed-length numeric sequence (typically of \num{512} bits) generated on the basis of the user attributes and a pair of keys, namely a \emph{master public key} (\textsl{mpk}) and a \emph{master key} (\textsl{mk}). In turn, \textsl{mpk} and \textsl{mk} are generated through a cryptographic parametric algebraic structure (e.g., a pairing group).
A message is encrypted via the \textsl{mpk} and the policy.
The users can decrypt the ciphertext by using the \textsl{mpk} and their own \textsl{sk}. It follows that an unauthorised user would not have the suitable \textsl{sk} as per the policy. Furthermore, without knowledge of the \textsl{mpk}, the user cannot read the encrypted data either. Notice that \textsl{mpk} alone would not allow for the generation of new \textsl{sk}s as the master key (\textsl{mk}) is also necessary. 
To conclude, we remark that the generation of keys, the encryption of plaintexts, and the deciphering thereof, are operations that are algorithmically handled and thus no trusted party is needed -- any peer with access to the required credentials could run the necessary code.

In our setting, intuitively, users are process participants, messages are the data artefacts exchanged during the process execution, ciphertexts are the encrypted data artefacts, policies determine which artefacts can be access by whom, and keys are the instruments that are granted to the process parties to try and access the artefacts.
Next, we explain how we combine the use of blockchain and CP-\ABE to build an access-control architecture for data exchanges in blockchains that meet the requirements listed in \cref{tab:requirements}.

%% file: sections/approach.tex
\begin{figure}[tbp]
    \includegraphics[width=\textwidth]{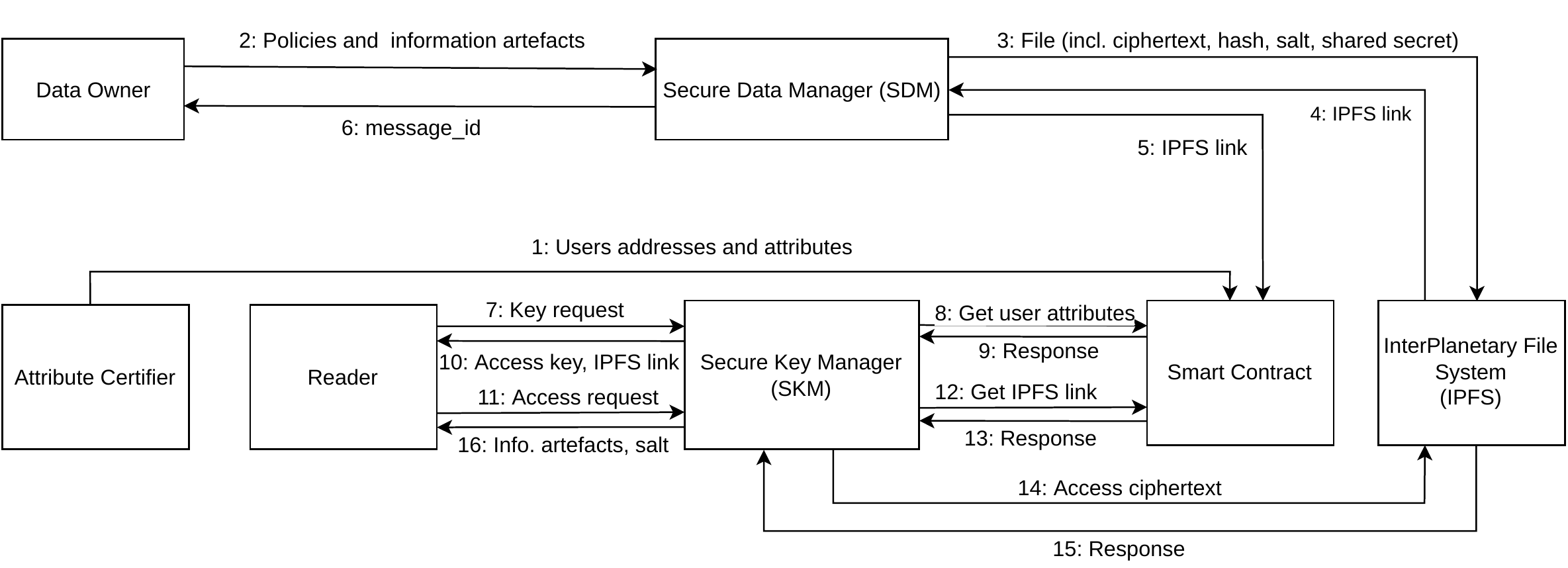}
    \caption{The key component interactions in the CAKE approach}
    \label{fig:sdm-skm}
\end{figure}

\def\CompA{Data Owner\xspace}
\def\CompB{Attribute Certifier\xspace}
\def\User{Reader\xspace}
\def\SC{Smart Contract\xspace}
In this section, we describe our approach, named Control Access via Key Encryption (CAKE). 
\Cref{fig:sdm-skm} illustrates the main components of our architecture alongside their interaction by means of a UML collaboration diagram. 
The involved parties are:
\begin{compactenum}
	\item the \CompA, who wants to cipher the information artefacts (henceforth also collectively referred to as \textit{plaintext}) with a specific access policy (e.g., the \textit{Manufacturer} who wants to restrict access to the bill of materials to the sole intended parties, i.e., the suppliers); we assume \CompA is equipped with a public/private key pair;
	\item one or more {\User}s, interested in some of the information artefacts (e.g., the \textit{Manufacturer}, the \textit{Electronic parts supplier}, the \textit{Mechanical parts supplier}); we assume every \User to keep their own public/private key pair;
	\item the \CompB, specifying the attributes characterising the potential readers of the information artefacts; we assume the \CompB to hold a blockchain account;
	\item the Secure Data Manager (SDM), a stateless software component ciphering the plaintext with the policy received from \CompA; we assume the \CompA to hold a blockchain account;
	\item the Secure Key Manager (SKM), a stateless software component generating access keys for {\User}s and that the {\User}s invoke to decrypt messages; we assume the SKM comes endowed with a pair of public and private encryption keys and to hold a blockchain account;
	\item \IPFS, used to store the ciphertext (i.e., the ciphered plaintext); and finally
	\item the \SC, used to safely store the resource locator to the ciphertext saved on \IPFS and the information about potential readers of the information artefacts; at deployment time, the \SC is associated with the blockchain account addresses of the SDM, of the SKM, and of the \CompB, so as to accept invocations only by those components.
\end{compactenum}

\noindent
Using the enumeration schema of~\cref{fig:sdm-skm}, action (1)
is a preliminary operation in which the \CompB transmits the attributes and the identifying blockchain account addresses of the {\User}s to the \SC so as to make them publicly verifiable on chain.
To this end, the \CompB operates as a push-inbound oracle~\cite{DBLP:conf/bpm/MuhlbergerBFCWW20}.
The \CompB stores on chain the attributes that determine the role of the \User and, optionally, the list of process instances in which they are involved.
For example, the \CompB stores on chain that \texttt{0x906D[\ldots]Dba8} is the address of a user that holds the \texttt{Manufacturer} attribute, determining the role, alongside the numeric identifier \texttt{14548487} for the running process (the so-called \emph{case id}), specifying the participation of the manufacturer in that particular process instance. Also, it registers that \texttt{0xE756[\ldots]b927} and \texttt{0xE2C8[\ldots]A2810} are {\User}s endowed with the \texttt{Supplier} and \texttt{14548487} attributes, and that the \texttt{Electronics} and \texttt{Mechanics} attributes belong to the first and the second \User, respectively.

Thereafter (2), the \CompA sends the plaintext (i.e., an information artefact such as the bill of material) and the access policies to the SDM, so that the latter can make use of the \ABE algorithm to cipher the plaintext with the policy.
The access policy declares the conditions according to which a user can be granted access to the ciphered information. 

Notice that a message can be separated into multiple \emph{slices}, and each of those can be associated to a different policy. For example, the bill of materials of process instance \texttt{14548487} is partitioned as follows (see \cref{tab:messagePolicies}): a slice is accessible to all suppliers and manufacturers involved in the process instance, as the policy reads \texttt{14548487 and (Manufacturer or Supplier)}; another one pertains to the sole production order of mechanical parts -- i.e., \texttt{14548487 and (Manufacturer or (Supplier and Mechanics))}; a third slice is specific for the electronic parts supplier -- i.e., \texttt{14548487 and (Manufacturer or (Supplier and Electronics))}. 
Notice that actors who are granted access to the data do not necessarily need to be directly involved in a process instance. It is the case of \textit{Customs}, e.g., in our running example: the policy reads, indeed, \texttt{Customs or (14548487 and ([\ldots]))}.
Therefore, \textit{Customs} are authorised to access data across all instances with their key, unlike \textit{Manufacturer}s.
In other words, the inclusion of a \texttt{case\_id} as an attribute in the policy determines a design choice on whether a \User can use the \textsl{sk} across different process instances or not. If the \texttt{case\_id} is specified, different access keys are generated for separate instances.

We assume every slice to be associated with a unique identifier (henceforth, \texttt{slice\_id}). \Cref{tab:messagePolicies} lists the policies used in our running example.
The semantics of access policies meets \textbf{R1.1}, as they are at the fine-grain level of slices within messages.

\begin{table}[tbp]
	\resizebox{.8\columnwidth}{!}{%
		\input{tables/message-policies}
	}
	\caption{Message policy examples}
	\label{tab:messagePolicies}
\end{table} 

Then (3), the SDM runs the algorithm for the generation of the ABE master public key (\textsl{mpk}) and master key (\textsl{mk}).
It uses the master key (\textsl{mk}) and the policies to encrypt the plaintext and attain the ciphertext.
Thereafter, it generates a unique identifier for the message (\texttt{message\_id}), such as \texttt{17071949511205323542}. For every slice, it builds a unique identifier (\texttt{slice\_id}) and a random number (named \emph{salt}) to be additively used for hashing.
Finally, it stores on IPFS the \texttt{message\_id} and, for each slice, the \texttt{slice\_id}, ciphertexts, hash of the slice's plaintext combined with the corresponding salt, and the following data encrypted with the public key of the SKM, which we collectively refer to as \emph{shared secret}:
\begin{iiilist}
	\item the \textsl{mpk},
	\item the \textsl{mk},
	\item additional parametric metadata for the cryptographic algebraic structure (for every slice).
\end{iiilist}
In our approach, the SDM forgets both \textsl{mpk} and \textsl{mk} after storing them as it is stateless.
Also, notice that a new pair of keys is created for every message (i.e., IPFS file) to address \textbf{R1.2}.
As a result (4), the IPFS returns the resource locator (i.e., the link to the IPFS file) to the SDM, which the SDM stores in the \SC (5).
Next (6), the SDM returns the \texttt{message\_id} to the \CompA.
The \CompA can send the \texttt{message\_id} to the interested parties to let them know the content is ready for retrieval.
For example, the \textit{Manufacturer} sends the suppliers the information that \texttt{17071949511205323542} is the identifier to use to fetch the bill of material.

As said, the SDM stores the association between the message and the resource locator on chain via the \SC (5).
Thus, we have data stored off-chain that is linked with the blockchain ledger, as per \textbf{R2}.
\begin{table}[tbp]
	\resizebox{\columnwidth}{!}{%
		\input{tables/message-encoding}
	}
	\caption{Examples of messages encoded by the CAKE system.}
	\label{tab:messageEncoding}
\end{table} 
\Cref{tab:messageEncoding} illustrates the messages we described in our running example in~\cref{sec:example} as saved on IPFS by the SDM.
Every IPFS file in our approach consists of a header with the address of the sender (i.e., the \CompA), the \texttt{message\_id}, and the encrypted pair of keys (\textsl{mpk} and \textsl{mk}). The body consists of slices, each with its identifier (\texttt{slice\_id}), hash, ciphertext, salt and metadata.
We recall that salt and metadata are encrypted with the public key of the SKM.
%
Furthermore, notice that the plaintext is encrypted, and albeit being stored semi-publicly on IPFS, it is unreadable even to the \CompA (unless a party obtains a suitable key, which can be granted only by the SKM). 
Thereby, we meet \textbf{R1.3}.

When 
the \User (e.g., the \textit{Electronic parts supplier}) wants to read the data of a message (e.g., the section of interest in the bill of materials), 
it requests a key from the SKM (7). 
Then, the SKM retrieves the \User data (the blockchain address and attributes) from the \SC (8,9). Notice that these pieces of information were previously stored by the \CompB at step (1). 
Equipped with these pieces of information and with the shared secret (including the \texttt{pk} and \texttt{mk}), it produces an ABE secret key (\textsl{sk}) for the \User and sends it back (10) together with the IPFS link corresponding to the requested message (e.g., the one identified by \texttt{17071949511205323542}).
Notice that the shared secret (including the \textsl{mpk} and \textsl{mk}) is saved on IPFS encrypted with the public key of the SKM, so that only the SKM can use its private key to retrieve the necessary information and produce the \textsl{sk}.
Also, we remark that the SKM is stateless, so it retains no information after it responds to the \User.

Equipped with their own access key (\textsl{sk}), the \User can begin the message decryption procedure.
As per the ABE paradigm, the \textsl{sk} alone is not sufficient to decipher messages though.
The \textsl{mpk} is also necessary, though it is encrypted in the IPFS file with the public key of the SKM.
Therefore, the \User makes an access request to the SKM (11). In turn, the SKM asks for the IPFS link from the \SC (12,13).
Then, the SKM retrieves the ciphertext from IPFS (14) and decrypts it with the \textsl{sk} of the user and 
the shared secret, extracted and deciphered from the requested message. 
If the decryption is successful, the SKM component sends the information artefacts back (15). 
Otherwise the \User request is denied.

Recall that a message can be composed of multiple slices.
In the case of the bill of material, e.g., message \texttt{17071949511205323542} consists of three slices (see \cref{tab:messagePolicies,tab:messageEncoding}).
The first slice contains information available to all suppliers, 
the second one only for \textit{Electronic parts supplier}, and the third one only for \textit{Mechanic parts supplier}.
Therefore, with the \textsl{sk} of the \textit{Electronic parts supplier}, its attributes and the shared secret kept by the SKM, the latter can decipher only the first and second slice, but not the third one -- as per the specified policies.
The SKM component thus returns those slices only (16).
The controlled, fine-grained data access in CAKE is designed to meet the requirements regarding auditability (\textbf{R3}), integrity and control (\textbf{R1}) and specifically granularity (\textbf{R1.1}).

We conclude this section with a few more remarks about security and integrity.
When a \User has received the information artefacts, they 
may want to verify that the data is not counterfeit.
This is the reason why the SKM component returns the (decrypted) \texttt{salt} along with the information artefacts to the \User (16). 
With the received deciphered data and the salt, the \User 
can compute the hash and check if it is equal to the one stored on IPFS by the SDM at step (3) or not. We remark that the \User had received the IPFS link along with the key at step (8), so that they could directly access the data on IPFS to check the integrity of the information artefacts received from the SKM later on.
This design contributes to meeting \textbf{R1.3}. The data on IPFS is ciphered and 
only the SKM can decipher it.
The usage of the salt prevents leakage of information, like
dictionary attacks.

Also, we remark that the communication backbone outside of blockchain and IPFS for the information exchanges between components is based on the \gls{ssl} protocol, so as to avoid packet sniffing from malicious third parties that could intercept the data.
Furthermore, we assume that the communication from \CompA to SDM, and from \User to SKM, are preceded by an initial authentication phase to address \textbf{R1.2}.
During a preliminary handshake, the SDM and the SKM send a random value to the callers. The callers responds with that value signed with their own private key, so as to let the invoked components verify their identity.
Notice that, without this measure, any malicious peer could request the \textsl{sk} in place of the real \User by knowing their address and guessing a file they could be granted access to.

%% file: tables/message-policies.tex
\begin{tabular}{|c|c|c|}
\hline
Message & Slice &  Policy \\ \hline
\begin{tabular}[c]{@{}c@{}}Purchase \\ order\end{tabular} & 1 &
\begin{lstlisting}[firstline=1,lastline=1]
14548487 and (Customer or Manufacturer)
\end{lstlisting}\\ 
\hline
 & 1 &
\begin{lstlisting}[firstline=2,lastline=2]
14548487 and (Manufacturer or (Supplier))
\end{lstlisting}\\ \cline{2-3}
 Bill of materials & 2 &
\begin{lstlisting}[firstline=3,lastline=3]
14548487 and (Manufacturer or (Supplier and Electronics))
\end{lstlisting}\\ \cline{2-3} 
 & 3 &
\begin{lstlisting}[firstline=4,lastline=4]
14548487 and (Manufacturer or (Supplier and Mechanics))
\end{lstlisting} \\
 \hline
\begin{tabular}[c]{@{}c@{}}Customs \\ clearance\end{tabular} & 1 &
\begin{lstlisting}[firstline=5,lastline=5]
Customs or (14548487 and (Manufacturer or (Supplier and Electronics)))
\end{lstlisting} \\ \hline
\begin{tabular}[c]{@{}c@{}}Invoice\end{tabular} &  1 &
\begin{lstlisting}[firstline=6,lastline=6]
14548487 and (Manufacturer or Client) 
\end{lstlisting}\\ \hline
\begin{tabular}[c]{@{}c@{}}Transportation \\ order\end{tabular} &  1 &
\begin{lstlisting}[firstline=7,lastline=7]
14548487 and (Manufacturer or Courier) 
\end{lstlisting} \\ \hline
\end{tabular}%


%% file: tables/message-encoding.tex
\begin{tabular}{|c|l|l|l|}
\hline
   Message
 & Original data
 & File header
 & File body (slices) \\ \hline
 
   \makecell{Purchase \\ order}
 & 
\begin{lstlisting}[firstline=1,lastline=5]
Company_name:            Alpha
Address:                 34, Alpha street
E-mail:                  cpny.alpha@mail.com
Quantity:                5
Price:                   $5000
\end{lstlisting}
 & 
\begin{lstlisting}[firstline=1,lastline=4]
sender:     0x989a [...] FeaD,
message_id: 2206302810394556865,
pk:         {"g": "\\u00af [...] 00f4},
mk:         {"beta": "\\u009b [...] 009a}
\end{lstlisting}
 & 
\begin{lstlisting}[firstline=1,lastline=5]
slice_id:   10322929677141064041,
hash:       0x2958 [...] fb611,
salt:       "\u008d [...] 01bd",
metadata:   {"c1": [...] 4Els},
cipherText: "qp21 [...] 7Ue9Q"
\end{lstlisting} \\ 
 \hline

 & 
\begin{lstlisting}[firstline=9,lastline=11]
Manufacturer_company:    Beta
Address:                 82, Beta street
E-mail:                  mnfctr.beta@mail.com
\end{lstlisting}
 &
 & 
\begin{lstlisting}[firstline=29,lastline=33]
slice_id:   7816105805828306901,
hash:       0x953a [...] f8d8,
salt:       "Zu00 [...] u004",
metadata:   {"c1": [...] 00a0},
cipherText: "oT2W [...] MQ=="
\end{lstlisting} \\ 
 \cline{2-2}\cline{4-4}
   \makecell{Bill \\ of \\ materials}
 & 
\begin{lstlisting}[firstline=13,lastline=18]
Frames_quantity:         8
Propeller_quantity:      80
PropellerGuard_quantity: 63
Camera_quantity:         30
Controller_quantity:     4
Amount_paid:             $12000
\end{lstlisting} 
 & 
\begin{lstlisting}[firstline=6,lastline=9]
sender:     0x906D [...] Dba8,	
message_id: 17071949511205323542,
pk:         {"g": "\\u0087 [...] 00ca},
mk:         {"beta": "\\u00b2 [...] 00fb}
\end{lstlisting} 
 & 
\begin{lstlisting}[firstline=35,lastline=39]
slice_id:   6847895862959863592,
hash:       0x12es [...] 1g23,
salt:       "bw32 [...] b464",
metadata:   {"c1": [...] asq2},
cipherText: "AS2w [...] btwd"
\end{lstlisting} \\ 
 \cline{2-2}\cline{4-4} 
 & 
\begin{lstlisting}[firstline=20,lastline=24]
IMU_quantity:            6
ESC_quantity:            40
Engines_quantity:        9
Batteries_quantity:      25
Amount_paid:             $9850
\end{lstlisting} 
 & 
 & 
\begin{lstlisting}[firstline=41,lastline=45]
slice_id:   3147899764966459866,
hash:       0xj4rs [...] ne3d,
salt:       "ns1w [...] mey4",
metadata:   {"c1": [...] 23rs},
cipherText: "ht3r [...] asf3"
\end{lstlisting} \\
 \hline
 
 \makecell{Customs \\ clearance}
 & 
\begin{lstlisting}[firstline=30,lastline=34]
Date:                    2022-05-10
Sender:                  Beta
Receiver:                Alpha

\end{lstlisting}
 & 
\begin{lstlisting}[firstline=11,lastline=14]
sender:     0x0182 [...] 8DC0,
message_id: 5757578887823057098,
pk:         {"g": "\\uKu00 [...] 00b9},
mk:         {"beta": "\\u004d [...] 0d2r}
\end{lstlisting}
 & 
\begin{lstlisting}[firstline=15,lastline=19]
slice_id:   4607011054135544290,
hash:       0x4ee6 [...] 2386,
salt:       \u0010 [...] 0013 ,
metadata:   {"c1": [...] 00c2},
cipherText: "udBA [...] IA=="
\end{lstlisting} \\
 \hline
 
 Invoice
 & 
\begin{lstlisting}[firstline=38,lastline=42]
Gross_total:             $5000
Company_VAT:             U12345678
Issue_date:              2022-05-12

\end{lstlisting}
 & 
\begin{lstlisting}[firstline=16,lastline=19]
sender:     0x906D [...] Dba8,
message_id: 6796003701952936428,
pk:         {"g": "\\u00dc [...] 00a2},
mk:         {"beta": "\\u00be [...] 00c0}
\end{lstlisting}
 & 
\begin{lstlisting}[firstline=22,lastline=26]
slice_id:   12641782614493395949,
hash:       0xad46 [...] 0f79,
salt:       "o9\u [...] 01e5,
metadata:   {"c1": [...] u09a},
cipherText: "7QsM [...] KVRS"
\end{lstlisting} \\ \hline
 
 \makecell{Transportation \\ order}
 & 
\begin{lstlisting}[firstline=46,lastline=49]
E-mail:                  cpny.alpha@mail.com
Arrival_date:            2022-06-06
\end{lstlisting}
 & 
\begin{lstlisting}[firstline=21,lastline=24]
sender:     0x906D [...] Dba8,
message_id: 9846697684368436866,
pk:         {"g": "\\ur25d [...] 3a7s},
mk:         {"beta": "\\u00lq [...] 08q2}
\end{lstlisting}
 & 
\begin{lstlisting}[firstline=8,lastline=12]
slice_id:   8655357017007860466,
hash:       0xe1de [...] 3f9f,
salt:       "\bvNA [...] 011n",
metadata:   {"c1": [...] 00b4},
cipherText: "opBK [...] J709"
\end{lstlisting} \\ \hline
\end{tabular}%

%% file: sections/imptes.tex
This section describes the proof-of-concept implementation of our approach and the test runs we conducted to assess its affordability for data access control and audits.

\begin{figure}[tbp]
	\includegraphics[width=\textwidth]{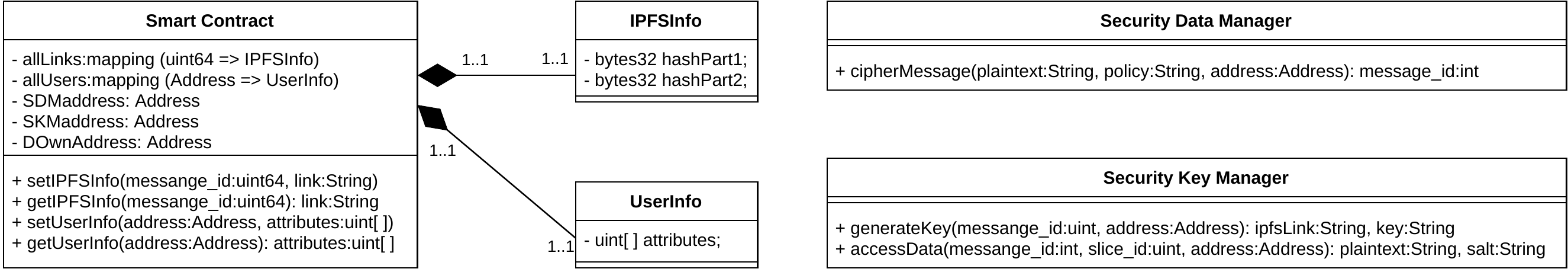}
	\caption{The implemented components of the CAKE system}
	\label{fig:uml}
\end{figure}
\Cref{fig:uml} depicts the core CAKE components in the form of a UML class diagram. The code of our prototype can be found at  \url{https://github.com/apwbs/AttributeBasedEncryption} together with the detailed results of our experiments. 
We implemented the SDM, SKM and the communication channels in Python.
We encoded the \SC in Solidity as we employ the Ethereum testnet Ropsten
for the deployment of our blockchain components:
all transactions directed to the CAKE \SC instance we used for our tests can be freely inspected at
\url{https://ropsten.etherscan.io/address/0x2D9EAe20E1E7515d47fBB9A5d454Ce7Be59cA03f}.
To manage the public/private key pair system for the \CompA, {\User}s and SKM, we resort to the Rivest–Shamir–Adleman (RSA) algorithm~\cite{DBLP:journals/cacm/RivestSA83}. 
In our software prototype, the length of the pair of keys amounts to \num{2048}~bits. 

To test our system, we called the methods of the deployed \SC to measure gas consumption.
More in the detail, we focussed on the invocations that require the payment of gas fees, namely
\begin{iiilist}
	\item the storage of the address and attributes of {\User}s (\texttt{setUserInfo(\ldots)} in \cref{fig:uml}), and
	\item the storage of the IPFS link associated to a message (\texttt{setIPFSInfo(\ldots)}).
\end{iiilist}
The data we used to run our experiments are taken from our running example (see \cref{tab:messageEncoding}).
We executed fifteen calls per day in five consecutive days. Out of the fifteen calls, ten were directed to \texttt{setUserInfo} and five to \texttt{setIPFSInfo}. 
The higher numerosity of the former is due to the fact that the latter has a rather fixed input format as the length in bits of IPFS locators is constant. As \texttt{setUserInfo} takes as input arrays, the variability of inputs is potentially higher.

\begin{table}[tbp]
	\resizebox{0.75\columnwidth}{!}{%
		\input{tables/experiments.tex}%
	}
	\caption{Gas consumption and total cost of test transactions}
	\label{tab:experiments}
\end{table} 
\Cref{tab:experiments} summarises the results. For every call we provide the average, minimum and maximum of
\begin{iiilist}
	\item units of gas used to run the code,
	\item price in wei paid for the gas consumption (using the Ropsten default setting), 
	\item total cost in Euros based on the daily exchange rate with an Ether. 
\end{iiilist}
The costs are relatively limited and range between ten and twenty Euro cents, which can be considered a reasonably low amount in light of the permanency and security guarantees provided by the system. Most importantly, the size of the information artefacts do not have a significant impact on the price paid to store them.

To save on gas expenditures, we have adopted a few mechanisms that reduced the size of input and output data. Among them, we recall the following two. First, we have turned IPFS links from their native base-\num{58} encoded format in strings of \num{46} characters to pairs of \texttt{bytes32} elements (the \texttt{IPFSInfo} struct in \cref{fig:uml}). This allowed for a saving of approximately \num{30000} units of gas per call of \texttt{setIPFSInfo}.
Secondly, we have encoded attributes into numeric identifiers to avoid the usage of strings for denumerable entities, thereby saving more gas units as the length of the attribute array increases.
Further gas-consumption optimisation techniques may be achievable especially for the attribute checking. This challenge paves the path for future work.

%% file: tables/experiments.tex
\begin{tabular}{@{} l S[table-format=5.2] S[table-format=10.0] S[table-format=1.5] S[table-format=5.0] S[table-format=10.0] S[table-format=1.5] S@{}}
	\toprule
	                   &               \multicolumn{3}{c}{\texttt{setIPFSInfo}}                &            \multicolumn{3}{c}{\texttt{setUserInfo}}             &                   \\ \cmidrule(lr){2-4}\cmidrule(lr){5-7}
	 & \textbf{gasUsed}    & \textbf{gasPrice}  & \textbf{Total cost} & \textbf{gasUsed}    & \textbf{gasPrice}  & \textbf{Total cost} & \textbf{ETH/EUR}  \\
	                   & \textbf{{[}unit{]}} & \textbf{{[}wei{]}} & \textbf{{[}EUR{]}}  & \textbf{{[}unit{]}} & \textbf{{[}wei{]}} & \textbf{{[}EUR{]}}  & \textbf{exchange} \\ \midrule
	avg.               & 67486.52            & 1399400015         & 0.16438             & 40755               & 1370810611         & 0.09734             & 1746.35           \\
	min                & 67484.6             & 1000000007         & 0.12378             & 40755               & 1000000007         & 0.07475             & 1650.68           \\
	max                & 67487               & 1649000034         & 0.18587             & 40755               & 1644053019         & 0.11191             & 1834.14           \\ \bottomrule
\end{tabular}

%% file: sections/sota.tex


Over the last few years, several research endeavours have been dedicated to the automation of collaborative processes based on blockchain.
Weber et al.~\cite{Weber.etal/BPM2016:UntrustedBusinessProcessMonitoringandExecutionUsingBlockchain}
present a technique that resorts to blockchain technology to execute business process between parties who do not trust each other. 
In their seminal work, they show how the actors can find a mutual agreement on the enacted behaviour 
without the need to trust a central authority for its enforcement. 
López Pintado et al.~\cite{Lopez-Pintado.etal/SPE2019:Caterpillar} 
present Caterpillar, 
a blockchain-based BPMN execution engine. Caterpillar allows users to create instances of a process and to monitor their status. 
Tran et al.~\cite{Tran.etal/BPMDemos2018:Lorikeet} introduce Lorikeet, a model-driven engineering (MDE) tool to implement business processes on chain for the management of assets (e.g., cars, houses), 
thereby proposing a solution for a scenario that traditionally relies on central authorities. 
Di Ciccio et al.~\cite{DiCiccio.etal/InfSpektrum2019:BlockchainSupportforCollaborativeBusinessProcesses} 
describe how to design and run business processes where several parties are involved, present the building blocks of model-driven approaches for blockchain-based collaborative business processes with a comparison between Caterpillar and Lorikeet. López Pintado et al.~\cite{Lopez-Pintado.etal/IS2022:ControlledFlexibilityBlockchainCollaborativeProcesses}  present a model to dynamically bind the actors in a multi-party business process to roles and a specification language for binding policies. 
CAKE can handle dynamic role binding as the attributes are set by the Attribute Certifier possibly at run time or deploy time. Access keys are generated upon request and not before the process starts.
Madsen et al.~\cite{Madsen.etal/FAB2018:CollaborationamongAdversaries:DistributedWorkflowExecutiononaBlockchain} 
investigate distributed declarative workflow execution where the collaboration is among adversaries. In such settings, the involved parties do not trust each other and they can also suspect that a party might not act like established. In this work, the authors demonstrate that the execution of the distributed declarative workflow could be implemented as a Smart Contract while ensuring the enforcement of workflow semantics and notarisation of the execution history. 
Corradini et al.~\cite{Corradini.etal/ACMTMIS2022:EngineeringChoreographyBlockchain} present ChorChain. It takes a BPMN choreography model as input and outputs its translation into a Solidity Smart Contract. 
ChorChain also allows auditors to obtain ex-post and runtime information 
on the process instances.
These works undoubtedly contribute to the integration of blockchain and process management thus unlocking security and traceability opportunities. However, they do not include mechanisms to ensure fine-grained access control to data saved on a public platform. In contrast, our work precisely focuses on this aspect in a collaborative business process scenario.

Another branch of research work that pertains to our investigation area is the privacy and integrity of data stored on chain. Several papers in the literature document the adoption of encryption to this extent. 
Hawk~\cite{Hawk} is a decentralised system that automatically implements cryptographic devices based on user-defined private Smart Contracts.
We take inspiration from this work in that we resort to policies backed by Smart Contracts to cipher messages.
Bin Li et al.~\cite{RZKPB} present RZKPB, a privacy protection system for transactions in shared economy built upon blockchain. This method does not require third trusted parties and preserves transaction privacy as it does not store the financial transactions publicly on chain. 
Their methodology relates with ours in that we resort to external data stores to save data too, yet we link it with transactions on the ledger.
In~\cite{FPPB}, the authors describe FPPB, a fast privacy protection method based on licenses. It uses zero-knowledge proof, secret address and encryption
primitives in the blockchain. Thanks to these features, it grants consistency without disclosing data. This architecture can be used in several shared economic applications. 
Rahulamathavan et al.~\cite{IoT-ABE} propose a new privacy-preserving blockchain architecture for IoT applications based on Attribute-Based Encryption techniques. We employ ABE too, yet with the objective of enhancing existing architectures with our approach. In contrast, this model aims at changing the blockchain protocol at its core. 
Benhamouda et al.~\cite{BenhamoudaCanAP} 
present a solution that allows a public blockchain to act as a repository of secret data. In their system, at first, a secret is stored on chain, then the conditions under which to release it are specified and, finally, the secret is disclosed if and only if the conditions are met. In our approach, we employ shared secrets among components but we do not use the blockchain as a storage for secret data nor expect to disclose the secret.
Differently from the techniques above, we tackle the problem of controlled data access in a multi-party process scenario, wherein several information artefacts are exchanged and different actors can access (parts of) messages based on fine-grained policies.

Wang et al.~\cite{EHR} present a secure electronic health record 
system wherein they combine ABE, Identity-Based Encryption (IBE) 
and Identity-Based Signature (IBS) 
with the blockchain technology.
This architecture differs from CAKE because in this case the hospital owns the data about patients, and patients specify the policies. 
In our case, no authority is intended to manage the data except the data owners themselves -- in healthcare processes, e.g., they would be the patients. 
Pournaghi et al.~\cite{MedSBA} provide a scheme based on blockchain technology and attribute-based encryption, named MedSBA. Their architecture differs from ours for two main reasons. Firstly, MedSBA makes use of two private blockchains, whereas we consider a public-blockchain scenario.
Secondly, they cipher the data with AES symmetric cryptography with a random key and then they cipher that random key via ABE. By ciphering with the AES encryption scheme, MedSBA does not allow different users to read the same message, or slices thereof.

%% file: sections/conclusion.tex
In this work, we have proposed CAKE, an approach that combines blockchain technology and \acrfull{abe} to control data access in the context of a multi-party business process. Our approach also makes use of IPFS to store information artefacts, access policies and meta-data. We employ Smart Contracts to store the user attributes, determining the access granted to the process actors, and the link to IPFS files. CAKE provides a fine-granular specification of access grants, data integrity, permanence and non-repudiability, allowing for auditability with minor overheads. 

An important aspect to analyse in future studies is the integration with alternative encryption methods.
For example, Odelu et al.~\cite{MobileABE} propose an RSA-based CP-ABE scheme with constant-size secret keys and ciphertexts (CSKC). Their approach 
targets high efficiency for limited-battery devices. 
The adoption of CSKC could be of help to integrate IoT devices in the management of blockchain-based processes.
Key-Policy Attribute-Based Encryption (KP-ABE)~\cite{KP-ABE} seems a promising asset for a more agile management of the process instance identifiers (case ids). 
With KP-ABE, attributes are associated with the ciphertext while the policy is associated with users, so the latter can decrypt the ciphertext only if the attributes of the encrypted text satisfy the user policy. 

We also plan to overcome existing limitations of our approach. If a \CompA wants to revoke access to data for a particular \User, e.g., they can change the policy and cipher the messages again. However, the old data on IPFS would still be accessible. 
To overcome this limitation, we are considering the usage of InterPlanetary Name System (IPNS), as it allows for the replacement of existing files, hence the substitution of a message with a new encryption thereof. 
Furthermore, we plan to turn the SDM and SKM into distributed components in order to make our architecture more robust. 
We are investigating the adoption of secure multi-party computation schemes~\cite{SMPC} to this end.

The integration of CAKE with existing blockchain-based process automation toolkits such as Caterpillar~\cite{Lopez-Pintado.etal/SPE2019:Caterpillar}, Lorikeet~\cite{Tran.etal/BPMDemos2018:Lorikeet} and ChorChain~\cite{Corradini.etal/ACMTMIS2022:EngineeringChoreographyBlockchain} is an interesting research avenue as well. 
CAKE can complement the control-flow-centric perspectives of the above tools with the data access control facilities it provides.
To this end, the automated translation of task-based authorisation constraints to policies would be part of the endeavour~\cite{DBLP:conf/bpm/WolterS07}.
Lorikeet specifically includes methods for on-chain data management, which CAKE can complement for confidential off-chain data.
As we resort to IPFS to store data, though, the integration should include oracles that permit Smart Contracts to interact with off-chain data~\cite{DBLP:conf/bpm/MuhlbergerBFCWW20,DBLP:conf/bpm/BasileGCK21}.
The system designer would then be able to determine the trade-off between full transparency on the decision process and access control, by balancing the on-chain and off-chain storage of data as discussed in~\cite{DBLP:conf/bpm/HaarmannBNW19}.
Finally, we aim to implement this system with other public blockchains in the future (e.g., Algorand~\cite{Chen.Micali/TCS2019:Algorand}) and test this system with real-world multi-party business processes in production.